\documentclass[10pt]{article}

\begin{document}

\title{Stellar models with Schwarzschild and non-Schwarzschild vacuum exteriors} 
\author{J. Ponce de Leon\thanks{E-mail: jpdel@ltp.upr.clu.edu; jpdel1@hotmail.com}\\ Laboratory of Theoretical Physics, Department of Physics\\ 
University of Puerto Rico, P.O. Box 23343, San Juan, \\ PR 00931, USA} 
\date{October  2007}

\maketitle

\begin{abstract}
A striking characteristic of non-Schwarzschild vacuum exteriors is that they contain not only the total gravitational mass of the source, but also an  {\it arbitrary} constant. In this work, we show that the constants appearing in the ``temporal Schwarzschild", ``spatial Schwarzschild" and  ``Reissner-Nordstr{\"o}m-like" exteriors are not arbitrary but are completely determined by  star's parameters, like the equation of state and the gravitational potential. Consequently, in the braneworld scenario  the gravitational field outside of a star is no longer determined by the total mass alone, but also depends on the details of the internal structure of the source. We show that the general relativistic upper bound on the gravitational potential $M/R < 4/9$, for perfect fluid stars, is significantly increased in these exteriors. Namely, $M/R < 1/2$, $M/R < 2/3$ and $M/R < 1$ for the temporal Schwarzschild, spatial Schwarzschild and  Reissner-Nordstr{\"o}m-like exteriors, respectively. We find that stellar models embedded in such exteriors are very diverse and rich in structure: For regular stars the deviation from the Schwarzschild exterior metric is automatically negligible, but in other limits they allow the existence of new kinds of stellar models, which have no general relativistic counterpart. Regarding the surface gravitational redshift, we find that the general relativistic Schwarzschild  exterior as well as the braneworld spatial Schwarzschild exterior lead to the  same upper bound, viz.,  $Z < 2$. 
However,  when the external
spacetime is the temporal Schwarzschild metric or the Reissner-Nordstr{\"o}m-like exterior 
there is no such constraint: $Z < \infty$. This infinite difference in the limiting
value of $Z$ is because for these exteriors the effective pressure at the surface is negative. 
The results of our work are potentially observable and can be used   to test the theory.

\end{abstract}

\medskip



\newpage

\section{Introduction}

In four-dimensional general relativity the spacetime outside of an isolated spherical star, without rotation,   is described by an unique line element, which is the Schwarzschild metric. The relevance of this is that the gravitational field outside of such a star depends solely on its  total gravitational mass.

Today, there are  several  theories that envision our world as embedded in a larger universe, with more than four dimensions \cite{Maartens1}-\cite{JPdeLgr-qc/0105120v2}. Although these theories have different purposes and motivations, they share a number of features \cite{JPdeLgr-qc/0111011}, and face the same challenge, namely the prediction of {\it observable} effects in $4D$ caused by new physics from  the extra dimensions. 

The study of the stellar structure, in the context of these theories, might constitute an important approach to reach this goal. 
However, there is a fundamental limitation. Namely, that there are many ways of producing, or embedding, a $4D$ spacetime in a given higher-dimensional manifold, while satisfying the field equations \cite{JPdeLgr-qc/0512067}. As a consequence,  the  effective
picture in four dimensions allows the existence of different possible non-Schwarzschild scenarios for the description of the spacetime outside of a spherical star.

A striking characteristic of non-Schwarzschild vacuum exteriors is that, besides the total gravitational mass, they  also contain an  {\it arbitrary} constant. Which suggests that the gravitational field outside of a star is no longer determined by the total mass alone. This gives rise to a number of questions,  e.g.,  what is the nature of  the constants in non-Schwarzschild exteriors?,  are they  absolute physical constants?, are they related to the source?

In this work we examine these questions in the context of various non-Schwarzschild vacuum exteriors  in five-dimensional braneworld theory, namely, the so-called ``temporal" and ``spatial"  Schwarzschild metrics as well as the ``Reissner-Nordstr{\"o}m-like" exteriors. With this aim we assume a simple model for the stellar interior and then use the matching conditions at the boundary to relate the internal and external metrics. We concentrate our attention on the following questions.

\medskip

(1) Can we relate the  constants in non-Schwarzschild exteriors to the structure of the source?

(2) On general grounds, can we restrict the value of these constants?

(3)  Do these exteriors impose some limit on the gravitational potential, similar to the general relativistic limit $M/R < 4/9$? 

(4) Do they change or eliminate the general relativistic upper bound  $Z  < 2$ on the  surface gravitational redshift?

\medskip

We are not going to discuss here the coupling of these metrics to the bulk geometry. Finding an exact solution in $5D$ that is consistent with a particular induced metric in $4D$ is not an easy task. However, the existence of such a solution is guaranteed by Campbell-Maagard's embedding theorems \cite{SanjeevWesson}, \cite{Indefenceof}. 

This paper is organized as follows. In section $2$ we choose a  model for the stellar interior. We adopt one that is similar to the general relativistic model for a sphere of homogeneous incompressible perfect fluid.
These models are specially suitable for our work because they are continously connected to the Schwarzschild interior solution, which serves as limiting case for a wide class of stellar models.  
Therefore, this choice  allows us to compare and contrast the properties of stars embedded in a non-Schwarzschild braneworld vacuum with some well-known properties of stars in ordinary general relativity.

In this framework, we will show here that the constants appearing in the above-mentioned non-Schwarzschild exteriors are {\it not} universal constants but change from one star to another, depending on the gravitational potential and the equation of state at the stellar core.

In section $3$ we consider the  temporal Schwarzschild vacuum exterior. The most interesting feature of this exterior is that, contrary to general relativistic perfect fluid models, the gravitational potential of a star  can get as near as one wants to $1/2$. That is,  there is {\it no} an upper limit on the surface gravitational redshift. This is a consequence of the non-local stresses induced in $4D$ from the Weyl tensor in $5D$, which allow the presence of negative (effective) pressures inside a star. Another important result is that the deviation from the Schwarzschild exterior becomes negligible for any star which, like our sun, is matter dominated and has a small gravitational potential. 

In section $4$ we consider the  spatial Schwarzschild exterior. We find two classes of stellar models. One of them is Schwarzschild-like in the sense that it works well for ``soft" equations of state at the core and small gravitational potentials. The other model, which we study in more detail,  is completely different because  it has no a Newtonian limit. In both models, the gravitational potential can approach $2/3$, but impose the same upper limit on the surface redshift as in general relativity.

In section $5$ we consider the Reissner-Nordstr{\"o}m-like exterior. In these models, the gravitational potential can approach $1$,  and there is no upper limit on the redshift of the light emitted from the surface of a star. Again this is allowed by the existence of negative pressure inside the source. In the weak field limit there is no difference with general relativity. 

In section $6$ we give a summary of our results. We emphasize that for ``regular" stars the deviation from the Schwarzschild exterior metric is automatically negligible: it is {\it not} an assumption. In addition, we point out that the  exteriors considered here allow the existence of two new kinds of stellar models, which have no general relativistic counterpart. One of them is what we call ``Quasi-Newtonian" stars, which are stars with small gravitational potential but whose internal pressure,  contrary to general relativity, is not negligible compared with the energy density.   The second kind of new  models are stars with negative pressure through the interior. The possible existence of such models is a consequence of the fact that the non-Schwarzschild exteriors under consideration allow negative  pressure at the boundary, while in general relativity the boundary  of an isolated star is a surface of zero pressure.

\section{Field equations on the brane}

The effective equations for gravity in $4D$ are obtained from dimensional reduction of the five-dimensional equations $^{(5)}G_{AB} = k_{(5)}^2 {^{(5)}T_{AB}}$. In particular,  in the  Randall $\&$ Sundrum  braneworld scenario \cite{Randall2}, where our universe is identified with a singular hypersurface (called {\it brane}) embedded in a   $5$-dimensional anti-de Sitter bulk $(^{(5)}T_{AB} = - \Lambda_{(5)}g_{AB})$ with ${\bf Z}_{2}$ symmetry  with respect to the brane, the effective equations in $4D$  are \cite{Shiromizu}
\begin{equation}
\label{EMT in brane theory}
^{(4)}G_{\mu\nu} = - {\Lambda}_{(4)}g_{\mu\nu} + 8\pi G T_{\mu\nu} +\epsilon k_{(5)}^4 \Pi_{\mu\nu} - \epsilon E_{\mu\nu},
\end{equation}
where $^{(4)}G_{\mu\nu}$ is the usual  Einstein tensor in $4D$; $\Lambda_{(4)}$ is the $4D$ cosmological constant, which is expressed in terms of the $5D$  cosmological constant $\Lambda_{(5)}$ and the brane tension $\lambda$, as 
\begin{equation}
\label{definition of lambda}
\Lambda_{(4)} = \frac{1}{2}k_{(5)}^2\left(\Lambda_{(5)} + \epsilon k_{(5)}^2\frac{ \lambda^2}{6}\right);
\end{equation}
$\epsilon$ is taken to be $- 1$ or $+ 1$, depending on whether the extra dimension is spacelike or timelike, respectively; $G$ is the Newtonian gravitational constant
\begin{equation}
\label{effective gravitational coupling}
8 \pi G = \epsilon k_{(5)}^4\frac{\lambda}{6};
\end{equation}
$T_{\mu\nu}$ is the energy momentum tensor (EMT) of matter confined in $4D$; $\Pi_{\mu\nu}$ is a tensor quadratic in $T_{\mu\nu}$ 
\begin{equation}
\label{quadratic corrections}
\Pi_{\mu\nu} = - \frac{1}{4} T_{\mu\alpha}T^{\alpha}_{\nu} + \frac{1}{12}T T_{\mu\nu} + \frac{1}{8}g_{\mu\nu}T_{\alpha\beta}T^{\alpha\beta} - \frac{1}{24}g_{\mu\nu}T^2;
\end{equation}
and $E_{\alpha \beta}$ is the projection onto the brane of the Weyl tensor in $5D$. Explicitly,  $E_{\alpha\beta} = {^{(5)}C}_{\alpha A \beta B}n^An^B$, where $n^{A}$ is the $5D$ unit vector $(n_{A}n^{A} = \epsilon)$ orthogonal to the brane. This quantity connects the physics in $4D$ with the geometry of the bulk.

Therefore, giving the EMT of matter in $4D$ is not enough to solve the above equations, because $E_{\alpha \beta}$ is unknown without specifying, {\it both} the metric in $5D$, and the way the $4D$ spacetime is identified \cite{JPdeLgr-qc/0511067}. In other words,  
the set of equations (\ref{EMT in brane theory}) is not closed in $4D$. The only quantity that can be specified without resorting to  the bulk metric, or the details of the embedding,  is the curvature scalar $^{(4)}R = {^{(4)}R}^{\alpha}_{\alpha}$,  because $E_{\mu\nu}$ is traceless. In particular, in empty space $(T_{\mu\nu} = 0, \Lambda_{(4)} = 0)$
\begin{equation}
\label{field eqs. for empty space}
^{(4)}R = 0.
\end{equation}

Solutions to this equation with spatial spherical symmetry,  have been discussed by a number of authors. In particular by Dadhich {\it et al} \cite{Dadhich}, Casadio {\it et al} \cite{Casadio}, Viser and Wiltshire \cite{Viser},  and Bronnikov {\it et al} \cite{Bronnikov}.

\subsection{Stellar interior}

An observer in $4D$, who is confined to making physical measurements  in our ordinary spacetime, can interpret the effective  equations (\ref{EMT in brane theory})  as the conventional Einstein equations with an effective EMT, 
$T_{\mu\nu}^{eff}$, defined as  
\begin{equation}
\label{def. of effective EMT}
8 \pi GT_{\mu\nu}^{eff} \equiv - {\Lambda}_{(4)}g_{\mu\nu} + 8\pi G T_{\mu\nu} + \frac{48 \pi G}{\lambda} \Pi_{\mu\nu} - \epsilon E_{\mu\nu}.
\end{equation}
In the case of a static, spherically symmetric distribution of matter in $4D$, with metric 
\begin{equation}
\label{metric in 4D}
ds^2 = e^{\nu(R)}dT^2 - e^{\sigma(R)}dR^2 - R^2(d\theta^2 + \sin^2\theta d\varphi^2),
\end{equation}
the effective density $\rho^{eff}$, radial pressure $p_{rad}^{eff}$ and tangential pressure $p_{\perp}^{eff}$ are expressed in terms of $\nu$ and $\lambda$ as follows
\begin{equation}
8\pi G\rho^{eff} = - e^{- \sigma}\left(\frac{1}{R^2} -  \frac{\sigma'}{R}\right) + \frac{1}{R^2},
\end{equation}
\begin{equation}
8\pi G p_{rad}^{eff} = e^{- \sigma}\left(\frac{\nu'}{R} + \frac{1}{R^2}\right) - \frac{1}{R^2},
\end{equation}
\begin{equation}
8\pi G p_{\perp}^{eff} = \frac{1}{2}e^{- \sigma}\left(\nu'' + \frac{\nu'^2}{2} + \frac{\nu' - \sigma'}{R} - \frac{\nu' \sigma'}{2}\right),
\end{equation}
where  a prime denotes derivative with respect to the radial coordinate $R$. 

We note that for a perfect fluid source with density $\rho$ and pressure $p$, from (\ref{def. of effective EMT}) the effective density and pressure are given by  $(\Lambda_{(4)} = 0)$
\begin{eqnarray}
\label{effective matter in terms of perfect fluid rho and p}
\rho^{eff} &=& \rho - \frac{\epsilon k_{(5)}^4}{48\pi G}\rho^2 - \frac{\epsilon E^{0}_{0}}{8\pi G}, \nonumber  \\
p^{eff}_{rad} &=& p - \frac{\epsilon k_{(5)}^4}{48 \pi G}(\rho + 2p)\rho + \frac{\epsilon  E_{1}^{1}}{8 \pi G}, \nonumber \\
p^{eff}_{\perp} &=& p - \frac{\epsilon k_{(5)}^4}{48 \pi G}(\rho + 2p)\rho + \frac{\epsilon  E_{2}^{2}}{8 \pi G}.
\end{eqnarray}
We will employ these expressions in our discussion in section $6$.

\subsection{Constant effective density}

In this work, we model the interior of a star by the solution of the effective field equations obtained 
under the assumptions 
\begin{equation}
\label{Shw. sol. for the effective equations}
\rho^{eff} = \rho_{0} = \mbox{constant},\;\;\;\;p_{rad}^{eff} = p_{\perp}^{eff}. 
\end{equation}
Thus, inside the star the metric will be the interior Schwarzschild metric \cite{Weinberg}, namely  
\begin{equation}
\label{uniform density interior in curvature coordinates}
ds^2 = \left(D - E\sqrt{1 - \frac{R^2}{a^2}}\right)^2 dT^2 - \left(1 - \frac{R^2}{a^2}\right)^{- 1}dR^2 - R^2(d\theta^2 + \sin^2\theta d\varphi^2),
\end{equation} 
where $D$, $E$ and $a $ are constants. This simple model, which in general relativity serves as a limiting 
case for {\it any} perfect fluid star\footnote{Is important to emphasize the role of isotropic pressures, because for anisotropic pressures there is no upper bound on the gravitational potential of a star. See \cite{Bowers}, \cite{Old JPdeL} and references therein.},  
will allow us to make contact with well known results in ordinary general relativity. The effective quantities are (in what follows we set $G = 1$),
\begin{equation}
\label{matter quantities}
8\pi  \rho^{eff} = \frac{3}{a^2}, \;\;\;8\pi  p^{eff} = \frac{3E\sqrt{1 - R^2/a^2} - D}{a^2(D - E\sqrt{1 - R^2/a^2})}.
\end{equation}
In the next three sections we will study in detail  the stellar models that result from the symbiosis between this stellar interior with the  braneworld exteriors mentioned in the introduction, namely, the temporal and spatial  Schwarzschild metrics as well the Reissner-Nordstr{\"o}m-like exteriors. 

\subsection{Schwarzschild exterior}

In order to facilitate the discussion, we briefly restate some results for the Schwarzschild exterior metric,
\begin{equation}
\label{Schw. exterior metric}
ds^2 = \left(1 - \frac{2M}{R}\right)dT^2 - \left(1 - \frac{2M}{R}\right)^{- 1}dR^2 + R^2(d\theta^2 + \sin^2\theta d\varphi^2).
\end{equation}
Instead of $E$ and $D$, it is useful to work with the quantity $\gamma$, which is the equation of state in the central region of the star. In the present case\footnote{To simplify the notation, in what follows we will suppress the ``eff'' over the matter quantities.}, 
\begin{equation}
\label{definition of gamma}
\gamma = \frac{p(0)}{\rho_{0}} = \frac{3 E - D}{3(D - E)}, \;\;\;\mbox{thus}\;\;\;D = \frac{3E(1 + \gamma)}{3\gamma + 1}.
\end{equation}
In general relativity, the boundary of an isolated star is a spherical surface of zero pressure. Using (\ref{matter quantities}) and (\ref{definition of gamma}) we find that the pressure becomes zero at $R = R_{b}$ satisfying $\sqrt{1 - R_{b}^2/a^2} = (1 + \gamma)/(3\gamma + 1)$. On the other hand, continuity of the metric requires $R_{b}^2/a^2 = 2M/R_{b}$. Thus, at the boundary surface
\begin{equation}
\label{phiSchw}
 \phi_{Schw} = \left(\frac{M}{R_{b}}\right)_{|Schw} = \frac{2\gamma(1 + 2\gamma)}{(1 + 3\gamma)^2}, \;\;\;\mbox{and} \;\;\;g_{TT}^{Schw}(R_{b}) = \left(\frac{1 + \gamma}{1 + 3\gamma}\right)^2.
\end{equation} 
For $\gamma \rightarrow \infty$ we recover the famous upper limit 
\begin{equation}
\label{Buchdahl limit for Schw sol}
\left(\frac{M}{R_{b}}\right)_{|Schw} = \frac{4}{9} \approx 0.444,\;\;\;g_{TT}^{Schw}(R_{b}) = \frac{1}{9} \approx 0.111,
\end{equation}
discovered by Buchdahl \cite{Buchdahl} for all static fluid spheres whose (i) energy density does not increase outward and (ii) the material of the sphere is locally isotropic. This limit sets an upper bound to the gravitational redshift of spectral lines from the surface of any star.

\section{Temporal Schwarzschild exterior}

In this section we consider in detail the model of a static spherical star, with  an interior described  by the line element (\ref{uniform density interior in curvature coordinates}), whose exterior spacetime is represented by a braneworld vacuum solution known as the temporal Schwarzschild  metric, viz., 

\begin{equation}
\label{temporal Schw exterior}
ds^2 = \left(1 - \frac{2{{M}}}{R}\right) dT^2 - \frac{(1 - 3{{M}}/2R)}{(1 - 2{{M}}/R)[1 - (3{{M}}/2R)\;c]}dR^2 - R^2 d\Omega^2,
\end{equation} 
where $c$ is an arbitrary dimensionless constant and $M$ is the total gravitational mass measured by an observer at spatial infinity\footnote{In order to avoid misunderstanding, we should immediately note that $M \neq (4\pi/3) \rho_{0} R_{b}^{3}$, except for the Schwarzschild exterior. More comments about this at the end of this section.} . For $c = 1$, this metric  reduces to the Schwarzschild exterior (\ref{Schw. exterior metric}).

The boundary conditions require continuity of the first and second fundamental forms at the surface of a star $R = R_{b}$.   For the case under consideration this amounts continuity of $g_{TT}$, $g_{RR}$ and $dg_{TT}/dR$ across $R_{b}$. 
Continuity of $g_{RR}$ gives
\begin{equation}
\label{continuity of grr}
\left(\frac{R_{b}}{a}\right)^2 = \frac{\phi \left[1 + 3c(1 - 2\phi)\right]}{2 - 3\phi}, \;\;\;\phi = \frac{M}{R_{b}}.
\end{equation}
Now, using(\ref{definition of gamma}), continuity of $g_{TT}$ yields
\begin{equation}
\label{continuity of gtt}
E = \sqrt{1 - 2\phi}\left[\frac{3(\gamma + 1)}{1 + 3\gamma} - \frac{\sqrt{1 - 2\phi}\sqrt{1 - 3c\phi/2}}{\sqrt{1 -  3\phi/2}}\right]^{- 1}.
\end{equation}
Finally, from the continuity of $dg_{TT}/dR$ we obtain 
\begin{equation}
\label{continuity of the derivative of gtt}
\phi = E\left(\frac{R_{b}}{a}\right)^2\frac{\sqrt{1 - 3\phi/2}}{\sqrt{1 - 3c \phi/2}},\;\;\;E > 0. 
\end{equation}
Combining (\ref{continuity of grr})-(\ref{continuity of the derivative of gtt}) we get a an expression for $\phi, \gamma$ and $c$ which can be solved to obtain $c$ as 
\begin{equation}
\label{equation for c}
c(\phi, \gamma) = \frac{3(q^2 - 12)\phi^2 - 2(q^2 - 15)\phi - 6 +  q(2 - 3\phi)\sqrt{4(1 - 3\phi)(1 - 2\phi) + q^2 \phi^2}}{6(1 - 2\phi)(1 - 3\phi)^2}, \;\;\; \mbox{with}\;\;\;\;\;\;q = \frac{3(\gamma + 1)}{1 + 3\gamma}.
\end{equation}
The sign in front of the root has been chosen positive to make sure that in the Schwarzschild limit $\phi \rightarrow \phi_{Schw}$, where $\phi_{Schw}$ is given by (\ref{phiSchw}), we recover 
 $c = 1$. In (\ref{equation for c}) the numerator as well as the denominator  tend to zero for $\phi \rightarrow 1/3$ and $\phi = 1/2$. Using L'hopital's rule, in this limit we find 
\begin{equation}
\label{c for phi = 1/3}
\lim_{\phi \rightarrow 1/3}{c} = \frac{5 + 6\gamma - 3\gamma^2}{3(\gamma + 1)^2}, \;\;\;\lim_{\phi \rightarrow 1/2}{c} = \frac{4}{3}.
\end{equation}
Since the quantity under the root in (\ref{equation for c}) is  positive, it follows that $c$ is a well defined quantity for {\it all} values of $\gamma$ and $\phi < 1/2$.

\subsection{Range of $c$}
We note that  $(\partial{c}/\partial\gamma)_{|\phi} < 0$, i.e., $c$ decreases monotonically with the increase of $\gamma$, for every fixed value of $\phi$. Therefore, the maximum value of $c$ is obtained for $\gamma = 0$, while its minimum is attained  in the limit $\gamma \rightarrow \infty$, namely\footnote{The function $c_{min} = c_{min}(\phi)$ is continuous at $\phi = 0.4$, but its first derivative is discontinuous; $(dc_{min}^{(1)}/d\phi) < 0$ while $(dc_{min}^{(2)}/d\phi) > 0$.} 
\begin{equation}
c_{min}^{(1)} = - \frac{1}{3(1-2\phi)}, \;\;\;\mbox{for}\;\;\;\phi \leq 0.4, \;\;\;\mbox{and}\;\;\;c_{min}^{(2)} = \frac{12 \phi - 5}{3(1 - 3\phi)^2}, \;\;\;\mbox{for}\;\;\;  0.4 \leq \phi < 1/2.
\end{equation}
Notice that $c_{min}^{(2)} = 1$ for $\phi = 4/9$. Thus, $c > 1$ for models with $\phi > 4/9$, for any  value of $\gamma$. In table $1$ we illustrate the range of $c$ for various values of $\phi$.
\begin{center}
\begin{tabular}{|c|c|c|c|c|c|c|c|c|c|c|} \hline
\multicolumn{11}{|c|}{\bf Table 1.   Range of $c$ for various values of $\phi$} \\ \hline
 \multicolumn {1}{|c|}{$\phi$} & 
$10^{- 6}$ &  $10^{-4}$ &$10^{- 2}$&$0.1$&$0.2$&$1/3$&$0.4$& $4/9$&$0.48$&\multicolumn{1}{|c|}{$0.499$} \\ \hline\hline
$c_{max} = c_{(\gamma = 0)}$ & $1.0000 $& $1.0003$& $1.02$&$1.23$&$1.50$ &$1.67$&$1.57$&$1.47$&$1.38$&$1.34$   \\ \hline
$c_{min} = c_{(\gamma \rightarrow \infty)}$ & $- 0.3333 $& $- 0.3334$& $- 0.34$&$- 0.42$&$- 0.56$ &$- 1.00$&$- 1.67$&$1.00$&$1.31$&$1.33$   \\ \hline
\end{tabular}
\end{center}
Table $1$ shows two things. Firstly, that for  objects that are more compact than $4/9$, the parameter $c$ is practically  insensitive to the 
change in $\gamma$, in the whole range $0 < \gamma < \infty$. In fact, for $\phi = 0.44$, $\phi = 0.49$ and $\phi = 0.499$ we find respectively 
\begin{equation}
\label{c for ultra compact stars}
c = 1.19 \pm 0.18, \;\;\;c = 1.17 \pm 0.08 \;\;\;\mbox{and } c = 1.334 \pm 0.001. 
\end{equation}
Secondly, that for less compact distributions  $c$ drastically changes with $\gamma$.  As an example, if we take $\phi = 1/3$ then $- 1 < c < 5/3 $, which is consistent with the limiting value found in (\ref{c for phi = 1/3}).

Substituting (\ref{equation for c}) into (\ref{continuity of grr}) and  (\ref{continuity of gtt}), we find that  $(R_{b}/a)$, $(1 - 3c\phi/2)$ and $E $ are positive and well defined for {\it all} values of $\gamma$,  $\phi < 1/2$ and $c$.  Thus, the  temporal Schwarzschild exterior allows, at least in principle, the existence of static spherical objects whose (geometrical) radius $R_{b}$
can be  very close to $2M$. The natural question to ask here is:  what is the physical mechanism that prevents the gravitational collapse of such compact stars? The answer to this question will be given in section $3.4$

\subsection{Braneworld exterior with $c \approx 1$}

In a model with $c \approx 1$, one would expect the gravitational potential $\phi$, as well as the equation of state,  of a braneworld star  not to be very much different from their  Schwarzschild values. Therefore, it makes sense to expand (\ref{equation for c}) in power series around $\phi_{Schw}$ and $\gamma_{Schw}$. To first order we obtain 
\begin{equation}
\label{expansion of c}
\delta c = \frac{2(1 + 3\gamma_{Schw})^2}{(\gamma_{Schw} + 1)^2}\delta \phi - \frac{4}{(1 + \gamma_{Schw})(1 + 3\gamma_{Schw})} \delta \gamma,  
\end{equation}
with\footnote{We note that for $\delta c = 0$, we get $\delta \phi/\delta \gamma = 2(\gamma_{Schw} + 1)/(1+ 3\gamma_{Schw})^3$, which coincides with the first derivative of (\ref{phiSchw}), as expected. }
\begin{equation}
\label{definitions of deltas for temp. Schw.}
\delta c \equiv (c - c_{Schw}) = (c -1),\;\;\;\delta \phi \equiv (\phi - \phi_{Schw}), \;\;\;\delta \gamma \equiv (\gamma - \gamma_{Schw}).
\end{equation}
This allows us to compare the parameters of a star in the Schwarzschild and the braneworld model under consideration.  

It is clear that a star, with a given gravitational potential, in a   braneworld model with $c > 1$  (or $c < 1$) will have a ``softer" (or ``stiffer") equation of state than in the Schwarzschild model. Similarly, for a given equation of state, the braneworld model with $c > 1$  (or $c < 1$) will produce a more (or less) compact  star than the Schwarzschild model.  The physical meaning of this will be discussed bellow in  section $3.4$.  

\subsection{Weak-field approximation}

In the weak-field limit the gravitational potential is very small. Thus, from (\ref{equation for c}) we get 
\begin{equation}
c = \frac{1 - \gamma}{1 + 3\gamma} + \frac{2(1 + 4\gamma - 3\gamma^2)}{(1 + 3\gamma)^2}\;\phi + O(\phi^2),\;\;\; \mbox{for}                        \;\;\;\phi \ll 1.
\end{equation}
Now we calculate the pressure inside the star.  Using (\ref{matter quantities}), (\ref{definition of gamma}) and (\ref{continuity of grr}), in this approximation we find 
\begin{equation}
\label{weak field limit for the temp. Schw. exterior}
p  = \gamma \rho_{0},
\end{equation}
{\it everywhere} inside the star. Notice that, in general,  the pressure is {\it not} negligible compared with  the energy density.

\paragraph{Newtonian stars:}
 In Newtonian astrophysics, as well as in general relativity, the boundary of an isolated star is a spherical surface of zero-pressure. Thus, from (\ref{weak field limit for the temp. Schw. exterior})  it follows that $\gamma = 0$ or $p = 0$ throughout the interior,  which in ordinary applications means $p \ll \rho$.

\paragraph{``Quasi-Newtonian" stars: } The temporal Schwarzschild exterior   does not demand zero-pressure at the boundary.  Therefore a small $\phi$ in this braneworld exterior,  does not necessarily require  $\gamma \rightarrow 0$   as in general relativity. Consequently, in principle the temporal Schwarzschild exterior allows the existence of quasi-Newtonian stars which  have a small gravitational potential, but are not necessarily dominated  by the energy density. This is a  pure consequence of the braneworld paradigm and has no counterpart in general relativity. More comments are given in section $6$.

\subsection{A closer look on stars with $ c > 1$ and $c < 1$}

Let us now discuss the pressure distribution inside the star. From (\ref{matter quantities}), and using (\ref{definition of gamma}) and (\ref{continuity of grr}), we obtain
\begin{equation}
\label{radial pressure for the temporal Schw. exterior}
8 \pi R_{b}^2 p(R) = \frac{3 \phi[1 + 3c(1 - 2\phi)]}{2 - 3\phi}\left\{\frac{\sqrt{1 - \frac{\phi[1 + 3c(1 - 2\phi)]}{2 - 3\phi}\left(\frac{R}{R_{b}}\right)^2}
 - \frac{\gamma + 1}{1 + 3\gamma}}{\frac{3(\gamma + 1)}{1 + 3\gamma} - \sqrt{1 - \frac{\phi[1 + 3c(1 - 2\phi)]}{2 - 3\phi}\left(\frac{R}{R_{b}}\right)^2}
}\right\}.
\end{equation}
In order to evaluate the pressure at the boundary, $R = R_{b}$, we solve (\ref{equation for c})  with respect to $q$:
\begin{equation}
\label{q}
q = \frac{3(\gamma + 1)}{1 + 3\gamma} = \frac{3\sqrt{1 - 2\phi}}{2\sqrt{1 - 3\phi/2}\sqrt{1 - 3c\phi/2}}[1 + c(1 - 3\phi)].
\end{equation}
(Note that for $\phi = 1/3$ we recover (\ref{c for phi = 1/3})). Then we substitute this expression into (\ref{radial pressure for the temporal Schw. exterior}) and after some algebra we get
\begin{equation}
\label{bound. pressure for temporal Schw}
8 \pi R_{b}^2 p(R_{b}) = \frac{3\phi (1 - c)}{2 - 3\phi}.
\end{equation}
This expression shows that the pressure at the boundary only becomes zero for $c = 1$, i.e., for the Schwarzschild exterior. In the braneworld model under study, it is positive for $c < 1$ and negative for $c > 1$.

In order to grasp this result, let us calculate the average slope of the pressure, 
\begin{equation}
\label{definition of average slope}
\frac{\Delta p}{\Delta R} = \frac{p(R_{b}) - p (0)}{R_{b} - 0}.
\end{equation}
Since, in the Tolman-Oppenheimer-Volkov equation of hydrostatic equilibrium, the  gravitational attraction is balanced by the pressure gradient 
$- dp/dR$, it follows that the magnitude of (\ref{definition of average slope}) can be interpreted as the ``average hydrostatic force" against gravitational collapse. Substituting the above expressions we find
\begin{equation}
\label{average slope of p for temporal Schw. exterior}
8\pi R_{b}^3\frac{\Delta p}{\Delta R} = - 6\gamma \phi - \frac{3\phi[1 + 3\gamma (1 - 2\phi)]}{2 - 3\phi}(c -1).
\end{equation}
It is evident that  for  $c > 1$ the (average) hydrostatic force is larger that in the case where $c \leq 0$.

As an illustration of this, let us consider a star with an  ultra-relativistic core, i.e., $\gamma = 1/3$. In the Schwarzschild model $(c = 1)$  this star has $\phi_{Schw} = 5/18 \approx 0.278$. Thus, 
\begin{equation}
\label{slope of pressure, Schw. case}
8 \pi R_{b}^3 \left(\frac{\Delta p}{\Delta R}\right)_{Schw} = - \frac{5}{9}\approx - 0.556. 
\end{equation}
The same star in the braneworld model under consideration with $c \neq 1$, but very close to $1$,  would have different  gravitational potential.  In this case, the deviation from Schwarzschild, to first order  in $\delta \phi$ and $\delta c$ from (\ref{definitions of deltas for temp. Schw.}) is given by  
\begin{equation}
8 \pi R_{b}^3 \left(\frac{\Delta p}{\Delta R}\right) =  - \left(\frac{5}{9} + 2 \delta \phi + \frac{40}{63} \delta c\right). 
\end{equation}
On the other hand, from (\ref{expansion of c}) it follows that $\delta c = 4.5 \times \delta \phi $. Substituting above we obtain,

\begin{equation}
8 \pi R_{b}^3 \left(\frac{\Delta p}{\Delta R}\right) =  - \left(\frac{5}{9} + \frac{68}{63} \delta c\right). 
\end{equation}
The physical interpretation in this case is clear: a braneworld star  with $\delta c > 0$ is more compact $(\delta \phi > 0)$ that in the Schwarzschild model, and therefore a larger hydrostatic force is needed to keep the static equilibrium  of the star. 

\subsection{Effective matter quantities outside the star}

An observer in $4D$, who is not directly aware of the existence of an extra dimension, will interpret the metric (\ref{temporal Schw exterior}) as if it were governed by an effective EMT,  
\begin{eqnarray}
\label{effective matter for the temporal Schw metric}
8\pi R_{b}^2\rho^{eff}(R) &=& \frac{3\phi^2(c - 1)}{[2 - 3\phi(R_{b}/R)]^2}\left(\frac{R_{b}}{R}\right)^4, \nonumber \\
8\pi R_{b}^2 p_{rad}^{eff}(R) &=& \frac{3\phi (1 - c)}{[2 - 3\phi(R_{b}/R)]} \left(\frac{R_{b}}{R}\right)^3, \nonumber \\
8\pi R_{b}^2 p_{\perp}^{eff}(R) &=& \frac{3\phi (c - 1)[1 - \phi (R_{b}/R)]}{[2 - 3\phi (R_{b}/R)]^2}\left(\frac{R_{b}}{R}\right)^3.
\end{eqnarray}
It  satisfies the radiation-like equation of state
\begin{equation}
\rho^{eff} = p_{rad}^{eff} + 2 p_{\perp}^{eff},
\end{equation} 
which follows from the fact that $E_{\alpha \beta}$ is traceless. 
In addition, we see that the radial pressure is continuous across the boundary of the star. 

\bigskip

We would like to finish this section with the following comments:

\medskip 

(i) In general relativity, when the field is time-independent, the gravitational mass $M_{g}$ inside a $3D$ volume $V_{3}$,  is given by the Tolman-Whittaker formula. By virtue of the Einstein field equations, this formula can be expressed in terms of the metric coefficients. 
For the static metric (\ref{metric in 4D}) it becomes

\begin{equation}
\label{grav. mass in terms of the metric}
M_{g}(R) = \frac{1}{2}R^2 e^{(\nu - \sigma)/2}\nu',
\end{equation}
 Evaluating this expression for the metric (\ref{temporal Schw exterior})  we obtain 
 \begin{equation}
\label{Mg for the temporal Schw}
M_{g}(R) = {{M}}\left[\frac{1 - 3c({{M}}/2R)}{1 - (3{{M}}/2R)}\right]^{1/2}.
\end{equation}
Asymptotically, for $R \rightarrow \infty$; we get $M_{g}(\infty) = M$, i.e., the total gravitational mass is given by $M$. The same result can be obtained from the analysis of the geodesic equation in the low velocity approximation.

(ii) The total mass $m$ defined by 
\begin{equation}
\label{inertial mass}
m = \int{T_{0}^{0}dV_{3}},
\end{equation}
where the integration is over the {\it whole} space, is different from $M$, except in the case where $c = 1$. 
Indeed after a simple calculation, using (\ref{matter quantities}), (\ref{continuity of grr}) and (\ref{effective matter for the temporal Schw metric}), we obtain
\begin{equation}
\label{m for temporal Schw. metric}
m = \frac{M (1 + 3c)}{4}.
\end{equation} 
For $c= 1$, $m = M$. This is identical to $P^{0}$, the zero component of the four-momentum vector $P^{\mu}$ of the body, which is calculated from the asymptotic behavior of the spatial part of the metric. Therefore $m = P^{0} = M(3 + c)/4$ can be identified with the ``inertial mass" of the central body and its gravitational fields.

(iii) As a consequence of the boundary conditions (\ref{continuity of grr}), (\ref{continuity of gtt}) and (\ref{continuity of the derivative of gtt}), both the Tolman- Whittaker mass as well as  the mass function $m(R) = (1 - e^{- \sigma(R)})R/2$ are continuous across the boundary of the star.

\section{Spatial Schwarzschild exterior}

In this section we study the model where the  spacetime surrounding an isolated star, which we describe  by the line element (\ref{uniform density interior in curvature coordinates}), is represented by a braneworld vacuum solution known as the spatial Schwarzschild metric, viz.,

\begin{equation}
\label{spatialsSchw exterior}
ds^2 = \frac{1}{b^2}\left(b - 1 + \sqrt{1 - \frac{2 b M}{R}}\right)^2\;dT^2 - \left( 1 - \frac{2 b M}{R}\right)^{- 1}dR^2 - R^2 d\Omega^2,
\end{equation}
where $M$ is the total gravitational mass measured at spatial infinity and $b$ is a dimensionless constant. For $b = 1$, we recover the Schwarzschild exterior metric.
 
Matching  conditions at the boundary $R = R_{b}$ require 
\begin{equation}
\label{positivity of mass}
M = \frac{E R_{b}^3}{a^2},
\end{equation}
and
\begin{equation}
\label{E and D}
E = \frac{1}{2b}, \;\;\; D = 1 - \frac{1}{b} + \frac{3}{2b}\sqrt{1 - \frac{2b M}{R_{b}}}.
\end{equation}
Since $M > 0$, from (\ref{positivity of mass}) it follows that $E > 0$. Consequently, the above equation indicates that $b > 0$ in this model. 

Now, substituting (\ref{definition of gamma}) into (\ref{E and D}) we obtain
\begin{equation}
\label{The root}
\sqrt{1 - \frac{2bM}{R_{b}}} = \frac{1 + \gamma - 2(b - 1)(1 + 3\gamma)/3}{(1 + 3\gamma)},
\end{equation}
from which we get the gravitational potential $M/R_{b}$, viz., 
\begin{equation}
\label{phi for the sol in curvature coord.}
\phi \equiv   \frac{M}{R_{b}} = \frac{2\left[1 + 2\gamma - (b - 1)(1 + 3\gamma)/3\right]\left[\gamma + (b- 1)(1 + 3\gamma)/3\right]}{b (1 + 3\gamma)^2}.
\end{equation}
The behavior of $\phi$ as a function of $\gamma$, for various values of $b$, is shown in Figure $1$. In the Buchdahl limit $\gamma \rightarrow \infty$ we find
\begin{equation}
\label{Buchdahl limit for spatial Schw. exterior}
\phi \rightarrow \frac{2}{3}\left(1 - \frac{b}{3}\right),
\end{equation}
which for  $b =1 $ reduces to  the Schwarzschild expression (\ref{phiSchw}).

\paragraph{Calculating $b$:}

When we solve (\ref{phi for the sol in curvature coord.}) with respect to $b$ we obtain two solutions, 
\begin{equation}
\label{b1}
b_{1} = \frac{9\gamma + 5}{2(3\gamma + 1)} - \frac{9}{4}\phi - \frac{3}{2}\sqrt{1 + \frac{9}{4}\phi^2 - \frac{(9\gamma + 5)}{(1 + 3\gamma)}\phi},
\end{equation}
and
\begin{equation}
\label{b2}
b_{2} = \frac{9\gamma + 5}{2(3\gamma + 1)} - \frac{9}{4}\phi + \frac{3}{2}\sqrt{1 + \frac{9}{4}\phi^2 - \frac{(9\gamma + 5)}{(1 + 3\gamma)}\phi},
\end{equation}
which correspond to  two different physical models\footnote{Both solutions coincide for $\gamma = \gamma_{min}$ (see (\ref{lower limit  on gamma})). In this case $b_{1} = b_{2} = (4 - 9\phi^2)/8\phi$. Note that $b_{1} = b_{2} \approx 1$ for $\phi \approx 0.357$, and $\gamma \approx 0.768$.}. The most evident difference is that models with $b = b_{1}$, as well as the Schwarzschild exterior of general relativity, are compatible with the Newtonian limit ($\gamma \approx 0$, $\phi \ll 1$). Indeed, in this limit $b_{1} = 1$ automatically. However, this is {\it not} the case for models with $b = b_{2}$, for which we get $b_{2} = 4$ in this limit. The behavior of these solutions is illustrated in Figures $2$ and $3$.

In order to get a better understanding of  the physics behind these two solutions,  let us set  $\phi = \phi_{Schw} $ in (\ref{b1}) and (\ref{b2}). We obtain,
\begin{equation}
b(\phi_{Schw}) = \frac{9\gamma^2 + 15\gamma + 5 \pm 3|1 + \gamma - 3\gamma^2|}{2(1 + 3\gamma)^2}. 
\end{equation}
Obviously we should get  $b(\phi_{Schw}) = 1$. The result depends on the sign of $(1 + \gamma - 3\gamma^2)$, namely,
\begin{eqnarray}
\label{The b = 1 condition}
b_{1}(\phi_{Schw}) &=& 1, \;\;\;\mbox{for}\;\;\;1 + \gamma - 3\gamma^2 > 0, \nonumber \\
b_{2}(\phi_{Schw}) &=& 1, \;\;\;\mbox{for}\;\;\;1 + \gamma - 3\gamma^2 < 0.
\end{eqnarray}
This indicates that  $b_{1}$ can take values close to $1$ only if\footnote{Here, $\gamma \approx 0.768$ is the approximate positive solution of
$(1 + \gamma - 3\gamma^2) = 0$. See the footnote $6$.} $\gamma < 0.768$. Similarly,  $b_{2}$ can take values close to $1$ only if $\gamma > 0.768$. 
From a physical point of view, this means that braneworld stars surrounded by  a  spatial Schwarzschild exterior with  $b = b_{1}$  should have  ``softer" equations of state than stars surrounded by an exterior with   $ b = b_{2}$. 

\paragraph{Degree of compactification:}

In order to avoid misunderstanding, we would like to emphasize that in the present model the compactness of a star is not given by $\phi$, but by the deviation of $g_{TT}(R_{b})$ from unity. To be more specific, let us consider the case where $b \approx 1$ and calculate $\phi$ and $g_{TT}(R_{b})$ to first order in $(b - 1)$. We obtain, 
\begin{equation}
\label{degree of compactification}
\phi = \phi_{Schw} + \frac{2(1 + \gamma - 3\gamma^2)}{3(1 + 3\gamma)^2}(b - 1), \;\;\;g_{TT}(R_{b}) = g_{TT}^{Schw}(R_{b}) - \frac{4(\gamma + 1)}{3(1 + 3\gamma)^2}(b - 1).
\end{equation}
This explain what we see in Figure $1$. Specifically,  for $b > 1$ we find that $\phi > \phi_{Schw}$ for ``low" values of $\gamma$ $(\gamma < 0.768)$, and $\phi < \phi_{Schw}$ for ``high" values of $\gamma$ $(\gamma > 0.768)$. However, $g_{TT}(R_{b}) < g_{TT}^{Schw}(R_{b})$, regardless of the equation of state at the center,  indicating that the star is more compact\footnote{$g_{TT}(R_{b})$ closer to $0$ corresponds to a star which is  much more ``compact" than a star with $g_{TT}(R_{b})$ closer to $1$.} $(b > 1)$. Exactly the opposite occurs for $b < 1$.

\medskip

 In summary, only the braneworld exterior with $b = b_{2}$ can be used to model compact stars with and equation of state stiffer than $\gamma = 0.768$ and  $b \approx 1$. In the remaining of this section we will study these models in more detail.

\subsection{Spatial Schwarzschild exterior with $b = b_{2}$}
Thus, if we know the gravitational potential of a given star, then the value of $b$ is determined by the equation of state at the center. However, contrary to models with\footnote{We note that $b_{1}$ as well as the metric coefficients are well defined in the whole range $\gamma_{min} \leq \gamma < \infty$ and $0 \leq \phi < 2/3$.} $b = b_{1}$, not every $\phi$ and $\gamma$ yield a reasonable model for $b = b_{2}$; they have to satisfy some general  conditions.

Firstly, the quantity under the root must be non-negative. Which yields a lower limit on $\gamma$, namely,  
\begin{equation}
\label{lower limit  on gamma}
\gamma \geq \gamma_{min}, \;\;\;\;\gamma_{min} = \frac{(2 - \phi)(9 \phi - 2)}{3(3\phi - 2)^2}, \;\;\;\phi < \frac{2}{3}.
\end{equation}

Secondly, we have to make sure that $\sqrt{1 - 2bM/R_{b}}$ is a positive quantity. Substituting (\ref{b2}) into (\ref{The root}) we obtain
\begin{equation}
\label{The root in terms of phi and gamma}
\sqrt{1 - \frac{2bM}{R_{b}}} = \frac{3}{2}\phi - \sqrt{1 + \frac{9}{4}\phi^2 - \frac{9\gamma + 5}{3\gamma + 1}\phi}.
\end{equation}
Positivity of this quantity requires 
\begin{equation}
\label{positivity of the root}
\phi > \frac{3\gamma + 1}{9\gamma + 5}.
\end{equation}
This inequality uncovers two important features of the models under consideration. 
\begin{enumerate}

\item It imposes a lower limit on $\phi$, viz.,
\begin{equation}
\label{lower limit on phi}
 \phi > \phi_{min}, \;\;\;\;\phi_{min} = 0.2,
\end{equation}
otherwise $\sqrt{1 - 2bM/R_{b}}$ is negative in the whole range $0 \leq \gamma < \infty$. From a physical point of view, this means that stellar models in  a  spatial Schwarzschild exterior with  $b = b_{2}$ have {\it no} Newtonian limit. 

\item It imposes an upper limit on $\gamma$ for $0.2 < \phi < 1/3$ Namely, 
\begin{equation}
\label{upper limit on gamma}
\gamma < \gamma_{max}, \;\;\; \gamma_{max} = \frac{5\phi - 1}{3(1 - 3\phi)},
\end{equation}
which is obtained by solving (\ref{positivity of the root}) with respect to $\gamma$ for $\phi < 1/3$. 
\end{enumerate}
Thus, 
\begin{equation}
\label{bounded above gamma}
\frac{(2 - \phi)(9 \phi - 2)}{3(3\phi - 2)^2} \leq \gamma < \frac{5\phi - 1}{3(1 - 3\phi)}, \;\;\;\mbox{for}\;\;\;0.2 < \phi < 1/3,
\end{equation}
and 
\begin{equation}
\label{unbounded above gamma}
\frac{(2 - \phi)(9 \phi - 2)}{3(3\phi - 2)^2} \leq \gamma < \infty, \;\;\;\mbox{for}\;\;\; 1/3 \leq \phi  < 2/3.
\end{equation}
From the above expressions we get the range of $b$ for every given value of $\phi$, viz.,
\begin{equation}
\label{range of b for gamma restricted above}
\frac{4 - 9\phi^2}{8\phi} < b < \frac{1}{2\phi},  \;\;\;\mbox{for}\;\;\;0.2 < \phi < 1/3.
\end{equation}
and 
\begin{equation}
\label{range of b for gamma not restricted above}
\frac{4 - 9\phi^2}{8\phi} < b < 3 - \frac{9}{2}\phi,  \;\;\;\mbox{for}\;\;\; 1/3 \leq  \phi < 2/3.
\end{equation}
We note that an increase in $\phi$ causes a decrease in the mean value of $b$, but increases its range\footnote{What we mean here is that if we represent $b$ as $b = \bar{b} \pm \Delta b$, with $\bar{b} = (b_{min} + b_{max})/2$ and $\Delta b = (b_{max} - \bar{b})$, then $\bar{b}$ decreases and $|\Delta b|/ \bar{b}$ increases with the increase of $\phi$.}.

Tables $2$ and $3$ illustrate these inequalities.
For example taking $\phi = 0.25$ we find $ 0.093 < \gamma < 1/3$. Then $b$ increases monotonically with $\gamma$ in the range $b \in (1.72, \;\;2)$. For this value of $\phi$, the change of $b$ is small compared to its mean value, therefore it can be expressed as $b = 1.86 \pm 0.14$. 

\begin{center}
\begin{tabular}{|c|c|c|c|c|} \hline
\multicolumn{5}{|c|}{\bf Table 2.  Upper bound on  $\gamma$ for $0.2 < \phi < 1/3$. Allowed values for $b$}\\ \hline
 $ \phi = 0.21$ & $- 0.035 < \gamma < 0.045 $& $ b = 2.26 \pm 0.12$& $g_{TT}(R_{b}) =  0.46 \rightarrow 0.34$& $g_{TT}^{Schw}(R_{b}) = 0.58$;  $\gamma_{Schw} = 0.19$   \\ \hline
$\phi = 0.25$ & $0.093 < \gamma < 1/3$& $b = 1.86 \pm 0.14$& $g_{TT}(R_{b}) =  0.40 \rightarrow 0.25$&$g_{TT}^{Schw}(R_{b}) = 0.50$; $\gamma_{Schw} = 0.26$    \\ \hline
$\phi = 0.29$ & $0.272 < \gamma < 1.154$& $ b = 1.56 \pm 0.16 $& $g_{TT}(R_{b}) =  0.36 \rightarrow 0.18$&$g_{TT}^{Schw}(R_{b}) = 0.42$; $\gamma_{Schw} = 0.37$    \\ \hline
\end{tabular}
\end{center}
In order to facilitate the comparison with regular general relativity, in the last column we provide the Schwarzschild values of $g_{TT}(R_{b})$, which we denote $g_{TT}^{Schw}(R_{b})$, for the surface gravitational potential given in the first column. We also calculate $\gamma_{Schw}$, the corresponding Schwarzschild equation of state at the center. 
Our notation $g_{TT}(R_{b}) = u \rightarrow v$ means that $g_{TT}$ at the boundary decreases monotonically  from $u$ to $v$ in the range of $\gamma$ and $b$ indicated in the table.
\begin{center}
\begin{tabular}{|c|c|c|c|c|} \hline
\multicolumn{5}{|c|}{\bf Table 3. Allowed values for $b$ in the range $1/3 \leq \phi \leq (- 4 + 2\sqrt{13})/9$} \\ \hline
 $ \phi = 1/3$ & $\gamma > 5/9 \approx 0.55$& $1.125 < b < 1.500 $& $g_{TT}(R_{b}) =  0.31 \rightarrow 1/9$&$g_{TT}^{Schw}(R_{b}) = 1/3$;  $\gamma_{Schw} = 0.58$ \\ \hline
$\phi = (- 4 + 2\sqrt{13})/9$ & $\gamma > 0.77$& $1.000 \leq  b < 1.398$& $g_{TT}(R_{b}) = 0.28 \rightarrow 1/9 $&$g_{TT}^{Schw}(R_{b}) = 0.29$;  $\gamma_{Schw} = 0.76$    \\ \hline
\end{tabular}
\end{center}

\subsubsection{Models with $b \geq 1$}

From (\ref{range of b for gamma not restricted above}) it follows that  
\begin{equation}
b \geq 1\;\;\;\mbox{for}\;\;\;0.2 < \phi \leq \frac{- 4 + 2\sqrt{13}}{9} \approx 0.357.
\end{equation}

Tables $2$ and  $3$ clearly illustrate  this inequality\footnote{The gravitational  potential $\phi = (- 4 + 2\sqrt{13})/9$ follows from the l.h.s. term in the inequality (\ref{range of b for gamma not restricted above}).}. They also evidence that for the same value of gravitational potential, a braneworld star (with $b > 1$) is more compact than in the Schwarzschild case and, in principle, it can have a {\it softer} equation of state at the core. For example, the first row of table $2$ shows that  $\gamma \approx 0$ for $\phi = 0.21$, which in usual general relativity requires $\gamma^{Schw} = 0.19$.

\subsubsection{Models with $b \approx 1$}

 In table $4$ we illustrate the approximate values of the parameters $\gamma$ and $b$ for $\phi$ in the range $(0.356, \;\;4/9)$ for which $b$ can be larger or less than $1$, depending on the equation of state at the center.  This is the only range for which $b$ can be very close to one. 

\begin{center}
\begin{tabular}{|c|c|c|c|c|} \hline
\multicolumn{5}{|c|}{\bf Table 4. $(- 4 + 2\sqrt{13})/9 < \phi \leq 4/9$: $b$ can be either $< 1$ or $> 1$}\\ \hline
 $ \phi = 0.36$ & $\gamma > 0.80$& $ 0.98 < b < 1.38$& $g_{TT}(R_{b}) =  0.28 \rightarrow 1/9$&$g_{TT}^{Schw}(R_{b}) = 0.28$;  $\gamma_{Schw} = 0.80$  \\ \hline
$\phi = 0.377$ & $\gamma > 1$& $0.90 < b < 1.30$& $g_{TT}(R_{b}) =  0.27 \rightarrow 1/9$&$g_{TT}^{Schw}(R_{b}) = 0.25$;  $\gamma_{Schw} = 1.03$   \\ \hline
$\phi = 4/9 \approx 0.444$ & $\gamma > 7/3 \approx 2.33$& $0.625 < b \leq  1.000$& $g_{TT}(R_{b}) =  0.2178 \rightarrow 1/9$&$g_{TT}^{Schw}(R_{b}) = 1/9$;  $\gamma_{Schw} = \infty$   \\ \hline
\end{tabular}
\end{center}

\subsubsection{Models with $b < 1$}
From (\ref{range of b for gamma not restricted above}) we find that 
\begin{equation}
b < 1, \;\;\;\mbox{for}\;\;\; \phi > \frac{4}{9},\;\;\;\mbox{and}\;\;\;b \rightarrow 0 \;\;\;\mbox{as}\;\;\; \phi \rightarrow \frac{2}{3}.
\end{equation}
The approximate values of the star parameters for $b < 1$ in the range for $4/9 < \phi < 2/3$ are exhibited in table $5$.

\begin{center}
\begin{tabular}{|c|c|c|c|c|} \hline
\multicolumn{5}{|c|}{\bf Table 5. $ 4/9 < \phi < 2/3$}\\ \hline
 $ \phi = 0.445$ & $\gamma > 2.35$& $ 0.623 < b < 0.998$& $g_{TT}(R_{b}) =  0.2174 \rightarrow 1/9$&No Schwarzschild counterpart    \\ \hline
$\phi = 0.500$ & $\gamma > 5.00$& $0.438 < b < 0.750$& $g_{TT}(R_{b}) =  0.18 \rightarrow 1/9$&No Schwarzschild counterpart   \\ \hline
$\phi = 0.600$ & $\gamma > 39.67$& $0.158 < b < 0.300$& $g_{TT}(R_{b}) =  0.14 \rightarrow 1/9$&No Schwarzschild counterpart   \\ \hline
\end{tabular}
\end{center}
These models are interesting, because they have  $\phi > 4/9$, and are {\it less} compact than the Schwarzschild solution of general relativity\footnote{ Although this might look atypical, it is completely in agreement with our discussion in (\ref{degree of compactification}).} in the Buchdahl limit for  which $\phi = 4/9$.
The case with $b < 1$ is important because the exterior metric becomes singular at $R_{h} = 2M/(2 - b)$, where $R_{h}$ defines the radius of the event horizon. However, from table $5$, it is clear that $R_{h}/R_{b} = 2\phi/(2 - b) < 1$.

\subsection{Buchdahl limit}

Tables $3$, $4$ and $5$ show that all models  with $1/3 \leq \phi < 2/3$, and different values of $b$, have the same Buchdahl limit, namely 
${g_{TT}(R_{b})} = 1/9 \approx 0.11$. 
Models with $0.2 < \phi < 1/3$ have no Buchdahl limit. 

It turns out that this is a general feature of stars with a spatial Schwarzschild exterior. Indeed, substituting (\ref{The root}) in (\ref{spatialsSchw exterior}), the metric coefficient $g_{TT}$ becomes\footnote{We note that $(dg_{TT}(R_{b})/db)_{|\gamma} < 0 $.}
\begin{equation}
\label{gtt for spatial Shw}
g_{TT}(R_{b}) = \frac{\left[\gamma + 1 + (b - 1)(1 + 3\gamma)/3\right]^2}{b^2(1 + 3\gamma)^2}.
\end{equation}
From which we obtain  
\begin{equation}
g_{TT} \rightarrow \frac{1}{9} \;\;\;\mbox{as}\;\;\; \gamma \rightarrow \infty,
\end{equation}
for any value of $b$. We note that this is the same limiting value found in the Schwarzschild model (\ref{Buchdahl limit for Schw sol}).

\subsection{Effective matter inside and outside the star}

Straightforward calculation, from (\ref{matter quantities}) gives 
\begin{equation}
8\pi \rho_{0} = \frac{6 b\phi}{R_{b}^2},
\end{equation}
and 
\begin{equation}
p(R) = \rho_{0} \frac{[(1 + 3\gamma)\sqrt{1 - 2 b\phi (R/R_{b})^2} - (\gamma + 1)]}{[3(1 + \gamma) - (1 + 3\gamma)\sqrt{1 - 2b \phi (R/R_{b})^2}]},
\end{equation}
inside the star. 
At the boundary $R = R_{b}$, using (\ref{The root})
\begin{equation}
p(R_{b}) = \rho_{0}\frac{(1 - b)(1 + 3\gamma)}{2 + b(1 + 3\gamma)}.
\end{equation}
The pressure vanishes at the boundary for $b = 1$, as expected. Braneworld stars with $b > 1$ $(b < 1)$ have negative (positive) pressure at the boundary. This is also expected, after the discussion in section $3.4$, because models with $b > 1$ $(b < 1)$ are more (less) compact than the Schwarzschild models.

\medskip

For completeness we provide the effective density and pressure outside the star
\begin{equation}
\rho^{eff} = 0,\;\;\;p_{rad}^{eff} = - 2p_{\perp}^{eff}, \;\;\; 8\pi R_{b}^2p_{\perp}^{eff} = \phi\; b (b - 1)\left(\frac{R_{b}}{R}\right)^3
\left[b - 1 + \sqrt{1 - 2b \phi \left(\frac{R_{b}}{R}\right)}\right]^{- 1}.
\end{equation}
It is not difficult to verify that the radial pressure is continuous across the boundary. We finish this section mentioning the relationship between the total gravitational mass and the inertial mass given by zero component of the four-momentum vector $P^{\mu}$,
\begin{equation}
m = P^{0} = b M.
\end{equation}

\section{Reissner-Nordstr{\"o}m-like exterior}

We now study the case where the exterior spacetime is described by the braneworld vacuum solution 

\begin{equation}
\label{Reis-Nords-like exterior}
ds^2 = \left(1 - \frac{2M}{R} + \frac{\eta M^2}{R^2}\right) dT^2 - \left(1 - \frac{2M}{R} + \frac{\eta M^2}{R^2}\right)^{- 1}dR^2 - R^2 d\Omega^2,
\end{equation} 
which for $\eta = 0$ reduces to the Schwarzschild exterior metric. From the continuity of the metric across the boundary surface, and using  (\ref{E and D}) we get
\begin{equation}
\label{E for RN exterior}
E = \frac{\sqrt{1 - 2\phi + \eta\phi^2}}{q - \sqrt{1 - 2\phi + \eta \phi^2}}, \;\;\;\mbox{with}\;\;\;q = \frac{3(\gamma + 1)}{1 + 3\gamma}.
\end{equation}
On the other hand, continuity of $dg_{TT}/dR$ yields
\begin{equation}
\label{on the other hand E for RN}
E =  \frac{1 - \eta \phi}{2 - \eta \phi}.
\end{equation}
Since $E > 0$, it follows that either $\eta \phi < 1$ or $\eta \phi > 2$. In order to keep contact with regular general relativity $(\eta = 0)$, we assume that
\begin{equation}
\label{condition on etaphi}
\eta \phi < 1,
\end{equation}
which ensures the positivity of mass\footnote{From the Tolman-Whittaker formula (\ref{grav. mass in terms of the metric}), it follows that $M > 0$ requires $(dg_{TT}/dR)$ positive everywhere. Then, from (\ref{uniform density interior in curvature coordinates}) we obtain the requirement $E > 0$.} 
In addition, because $(1 - 2\phi + \eta \phi^2) > 0$ we obtain
\begin{equation}
\label{range for eta}
\frac{2 \phi - 1}{\phi^2} < \eta < \frac{1}{\phi}.
\end{equation}
From (\ref{E for RN exterior})  and (\ref{on the other hand E for RN}) we obtain a cubic equation for $\eta$, which can be written as
\begin{equation}
\label{cubic equation}
4\phi^4 \eta^3 + \phi^2(4 - q^2 - 20 \phi) \eta^2 - \phi(12 - 2 q^2 -33\phi)\eta + (9 - q^2 - 18\phi) = 0.
\end{equation}
For every given $\phi$ and $\gamma$ this equation yields three possible values for $\eta$. The physical solution is the one that satisfies (\ref{range for eta}). Setting $\eta = 0$ we recover the Schwarzschild case. However,  if we set $\phi = \phi_{Schw}$  this equation gives three solutions: one of them is $\eta = 0$ and the other two are unphysical because $\eta \phi_{Schw} > 1$. 

In the Buchdahl limit $\gamma \rightarrow \infty$ we find a simple solution, 
\begin{equation}
\eta_{(\gamma \rightarrow \infty)} = \frac{12 \phi - 3 - \sqrt{9 - 8\phi}}{8\phi^2}. 
\end{equation}
Unfortunately, for any finite $\gamma$ the analytical expression for $\eta$, in terms of $\phi$,   is extremely  cumbersome.

In table $6$ we present the numerical (physical) solution of (\ref{cubic equation}) for $\phi = 1/2$ and various values of $\gamma$. It illustrates that, for every fixed value of $\phi$, both  $\eta$ and $g_{TT}(R_{b})$ decrease, with the increase of $\gamma$.

\begin{center}
\begin{tabular}{|c|c|c|c|c|c|c|c|c|} \hline
\multicolumn{9}{|c|}{\bf Table 6. $\eta$ for $\phi = 1/2$ and various values of $\gamma$} \\ \hline
 \multicolumn {1}{|c|}{$\gamma$} & 
$10^{- 6}$ &  $10^{-2}$ &$0.1$&$0.2$&$1/3$&$1$&$10$&\multicolumn{1}{|c|}{$\infty$} \\ \hline\hline
$\eta$ & $1.363$& $1.348$& $1.226$&$1.115$&$1.000$ &$0.713$&$0.424$&$0.382$ \\ \hline
$g_{TT}$ & $0.341$& $0.337$& $0.306$&$0.279$&$0.250$ &$0.178$&$0.106$&$0.095$ \\ \hline
\end{tabular}
\end{center}
We note that for $\phi < 4/9$, $\eta$ can be either positive or negative: $\eta > 0$ $(\eta < 0)$ for $\gamma < \gamma_{(\eta = 0)}$ $(\gamma > \gamma_{(\eta = 0)})$ given by 
\begin{equation}
\gamma_{(\eta = 0)} = \frac{3\phi - 1 + \sqrt{1 - 2\phi}}{4 - 9\phi}, \;\;\;\phi < \frac{4}{9}.
\end{equation}
For $\phi \geq 4/9$, $\eta$ is always positive.

\subsection{Models with $\eta \approx 0$}

For $|\eta| \ll 1$, the solution of (\ref{cubic equation}), to first order in $\eta$, can be expressed as
\begin{equation}
\phi = \phi_{Schw} + \frac{2\gamma(1 + 2\gamma)(7\gamma^2 + 5\gamma + 1)}{3(1 + 3\gamma)^4}\;\eta + O(\eta^2).
\end{equation}
Then, for $g_{TT}$ we obtain
\begin{equation}
g_{TT}(R_{b}) = g_{TT}^{Schw}(R_{b}) - \frac{4\gamma (1 + 2\gamma)(1 + \gamma)^2}{3(1 + 3\gamma)^4}\; \eta + O(\eta^2).
\end{equation}
Thus, a braneworld exterior  with $\eta > 0$ $(\eta < 0)$ yields more (less)
compact stars than in the Schwarzschild model.

In this case, the pressure at the boundary is given by
\begin{equation}
\label{boundary pressure for RN-like exterior}
p(R_{b}) = - \rho_{0} \frac{\eta \phi}{3(2 - \eta \phi)}.
\end{equation}
As expected, we find that $p(R_{b}) < 0$ $(p(R_{b}) > 0)$ for  $\eta > 0$ $(\eta < 0)$,  and $p(R_{b}) = 0$ for Schwarzschild.

\subsection {Weak field approximation}
In the weak field limit  $\phi \ll 1$, from (\ref{cubic equation}) it follows that $q \approx 3$. Therefore, in the case under consideration $\phi \rightarrow 0$ demands $\gamma \rightarrow 0$. However, the opposite is not true, i.e., $\gamma = 0 $ does not require small $\phi$. 

Thus, contrary to four-dimensional general relativity, braneworld models allow the existence of stars with zero, and even negative,  (effective) pressure at the center.  In order to grasp this, let us keep in mind that all what is  required by the equation of hydrostatic equilibrium is $dp/dR  < 0$. In general relativity,  the Schwarzschild exterior demands $p = 0$ at the boundary, which eliminates stars with negative (isotropic) pressure. However, braneworld stars  can have  negative pressure at the boundary, providing the negative slope for the pressure necessitated for hydrostatic equilibrium.

\subsection{Models with $\eta = 1$: Extremal Reissner-Nordstr{\"o}m-like exterior}
From (\ref{cubic equation}) it follows that  $\eta \rightarrow 1$ as $\phi \rightarrow 1$, for all values of $\gamma$. In this limit $g_{TT} \rightarrow 0$. However, for $\eta = 1$ the gravitational potential $\phi$ is not necessarily $1$; the exterior metric resembles the so-called 
``extremal Reissner-Nordstr{\"o}m" metric.

Setting $\eta = 1 $ in (\ref{Reis-Nords-like exterior}) we get
\begin{equation}
\label{extremal Reis-Nords-like exterior}
ds^2 = \left(1 - \frac{M}{R}\right)^2 dT^2 - \left(1 - \frac{M}{R}\right)^{- 2}dR^2 - R^2 d\Omega^2.
\end{equation}
The equation relating $\phi$ and $\gamma$ is obtained from (\ref{cubic equation}), viz.,
\begin{equation}
4\phi^4 - 20\phi^3 + (37 - q^2)\phi^2 - (30 - 2q^2)\phi -q^2 + 9 = 0.
\end{equation}
It has four solutions: $\phi_{1} = (3 + q)/2$; $\phi_{2} = (3 - q)/2$ and $\phi_{3} = \phi_{4} = 1$. The solution of interest to us is the one 
where $\phi < 1$ for all values of $\gamma$.  It is
\begin{equation}
\label{phi for extremal RN exterior}
\phi = \frac{3\gamma}{1 + 3\gamma}.
\end{equation}
Thus, $\phi \approx 0$ for $\gamma \approx 0$ and $\phi \rightarrow 1$ for $\gamma \rightarrow \infty$. 

For completeness we provide the exterior effective quantities
\begin{equation}
\rho^{eff} = - p_{rad}^{eff} = p_{\perp}^{eff}, \;\;\;8\pi R_{b}^2 \rho^{eff} = \eta \phi^2\left(\frac{R_{b}}{R}\right)^4,
\end{equation} 
and notice that for this model $m = M$.

\section{Summary and conclusions}

 The set of effective equations in $4D$ does not provide enough information for the complete specification of the geometry in the brane. In the vacuum region outside the surface of a star there is  one equation $^{(4)}R = 0$ for the two metric functions $g_{TT}$ and $g_{RR}$. This is a second order differential equation for $g_{TT}$ and first order for $g_{RR}$. Therefore, for any smooth function $g_{TT}$,  solving a first order differential equation we obtain $g_{RR}$ containing an arbitrary integration constant. 

As in any other branch of physics, here the constants of integration have to be specified from the initial data and/or boundary conditions. In this work, within the framework of models with uniform effective density, and using standard matching conditions at the boundary surface,  
we have found that  the gravitational potential and the equation of state completely determine the value of  the integration constants $c$, $b$ and $\eta$ which appear, respectively,  in the temporal Schwarzschild exterior, spatial Schwarzschild exterior and  Reissner-Nordstr{\"o}m-like exterior. 

Thus, we have shown that the gravitational field in the braneworld vacuum region outside the boundary surface {\it does} depend on the interior structure of a star through the constants $c$, $b$ and $\eta$. However, we have not discussed  here the question of how this would affect the motion of test particles. We will discuss this question elsewhere.

The consideration of models with uniform  effective density and isotropic effective pressures is motivated by the Schwarzschild interior solution,  which sets an upper limit to the gravitational redshift of spectral lines from the surface of {\it any} (perfect fluid) star.  Besides, their simplicity allows us  to obtain manageable analytical expressions.  Thus, such models reveal in a simple way the new features incorporated  by the deviation from the general relativistic Schwarzschild exterior.

Without specifying an equation of state for the effective matter we cannot calculate the actual value of these constants. Therefore,  we have obtained  the full range of   $c$, $b$ and $\eta$ in terms of the gravitational potential.  These are given in sections $(3.1)$, $(4.1)$ and equation (\ref{range for eta}), respectively.

For the sake of generality, we have discussed all possible physical scenarios  in the whole range of these parameters.      
We have found that stars embedded in  exteriors with $c > 1$, $b > 1$ and $\eta > 0$ are more compact than stars in  exteriors with $c \leq 1$, $b \leq 1$ and $\eta \leq 0$. 

We demonstrated that such stars must have negative effective pressure, at least, near  the boundary in order to provide the pressure gradient  required to counterbalance the inward pull of gravity.  If the extra dimension is spacelike $(\epsilon = -1)$, then the projected Weyl tensor $E_{\mu\nu}$ plays a crucial  role in the properties of a star. Indeed, assuming that the {\it hydrostatic} pressure $p$ is positive, then  from (\ref{effective matter in terms of perfect fluid rho and p}) it follows that a  
negative effective pressure inside the source is a direct  consequence of a large positive contribution from $E_{1}^{1}$. In addition, since $E_{0}^{0} = - 3 E_{1}^{1}$ a large positive 
$E_{1}^{1}$ could give rise to the phenomenon of gravitational repulsion.  

If the extra dimension is timelike\footnote{Based on our current knowledge, models having a large timelike extra dimension cannot be dismissed as mathematical curiosities in non-physical solutions. See \cite{JPdeLgr-qc/0212058} and references therein. } $(\epsilon = 1)$, the role of the Weyl tensor diminishes because the effective density and pressure can become negative  in very dense stages, when the quadratic terms in (\ref{effective matter in terms of perfect fluid rho and p}) become dominant, independently of the concrete contribution from the Weyl tensor.

It should be noted that the effective matter quantities do not have to satisfy the regular energy conditions \cite{Bronnikov2}, because they involve terms of  geometric origin. In particular,  the radial effective pressure outside a star can be positive, negative or zero. But this has a strong influence on the interior of the stellar model, because the continuity of the second fundamental form across the boundary requires continuity of the radial pressure. Thus, a negative geometrical radial pressure outside of the source requires a negative effective pressure inside the source, which in turn allows the existence of much more compact perfect fluid stars than in ordinary general relativity.

We have shown that  these exteriors impose an upper bound on the gravitational potential, which is larger than in general relativity. Namely,  

\begin{equation}
\label{upper limit on phi for all models}
\left(\frac{M}{R_{b}}\right)_{|Schw} < \frac{4}{9}, \;\;\;\;\;\left(\frac{M}{R_{b}}\right)_{|TSchw} < \frac{1}{2},\;\;\;\;\;\left(\frac{M}{R_{b}}\right)_{|SSchw} < \frac{2}{3}, \;\;\;\;\;\left(\frac{M}{R_{b}}\right)_{|RN} < 1,
\end{equation}
for the Schwarzschild, temporal Schwarzschild (TSchw), spatial Schwarzschild (SSchw) and Reissner-Nordstr{\"o}m-like (RN) exteriors, respectively. Besides, 
\begin{equation}
c > 1, \;\;\;b < 1,\;\;\;\eta > 0, \;\;\;\mbox{for stars with} \;\;\;\phi \geq 4/9.
\end{equation}

Our analysis shows that braneworld stars are very diverse and rich in structure. In particular, there are four types of limiting configurations.

\begin{enumerate}

\item Newtonian stars: All exteriors considered here, except the one with $b = b_{2}$, are compatible with the Newtonian limit, in the sense that for $\phi  \ll 0 $ and $\gamma \approx 0$ we automatically recover $c = 1$, $b_{1} = 1$ and $\eta = 0$. This is an important result because preserves the observationally tested predictions of general relativity. In this context, it is interesting to mention that in five-dimensional Kaluza-Klein gravity there is only one non-Schwarzschild exterior that is consistent with this limit \cite{JPdeLgr-qc/0701129v2}.

\item  ``Quasi-Newtonian" stars: For $\phi \ll 1$, we expand our solution and find that the effective pressure  becomes (in the original notation)
 $p^{eff} \approx \gamma \rho^{eff}$ throughout a star. 
Since the braneworld exteriors do not require zero pressure at the boundary, contrary to what happens in Newtonian stars, the effective pressure is not necessarily negligible compared to the effective energy density. Rather from (\ref{effective matter in terms of perfect fluid rho and p}), neglecting the quadratic terms,   we find
\begin{equation}
p \approx \gamma \rho  - \frac{\epsilon (1 - 3\gamma)E^{1}_{1}}{8\pi},
\end{equation}
for the hydrostatic pressure and density. 

\item Stars with negative effective pressure: Setting $\gamma = 0$ we obtain models with negative effective pressure throughout the source. It decreases from zero at the center to a negative value at the boundary, thus providing the negative slope for the pressure to counterbalance the gravitational attraction and prevent the collapse. Such models have no analog in general relativity. The only thing that can be said about the hydrostatic pressure and density is that $p < \rho/3$, which follows from 
\begin{equation}
\label{trace of the effective EMT}
\rho - 3p = (\rho^{eff} - 3p^{eff}) - \frac{ \epsilon k_{(5)}^4}{24 \pi}\rho (\rho + 3p), 
\end{equation}
for a spacelike extra dimension. 

\item ``Quasi-black holes": The  temporal Schwarzschild exterior allows, at least in principle, the existence of static spherical objects whose (geometrical) radius $R_{b}$
can be very close to $2M$. In fact,  in the Buchdahl  limit $\gamma \rightarrow \infty$ we find $\phi \rightarrow 1/2$, i.e., $g_{TT}(R_{b}) \rightarrow0$,  for $c \rightarrow 4/3$. Similarly, in the Reissner-Nordstr{\"o}m-like  exterior, with the appropriate choice of $\phi$ the quantity $g_{TT}(R_{b})$ can be as near as one wants to $0$, this is illustrated in table $6$. Such extremely compact objects are possible because the pressure is negative at the boundary, so that the magnitude of the gradient of pressure in these models is greater than in the Schwarzschild stars   
in the Buchdahl limit. Because of the extremely high surface gravitational red shift,  from an observational point of view these objects are indistinguishable from ``real" black holes.

\end{enumerate}

The analysis of the spatial Schwarzchild exterior yields two different models, for $b = b_{1}$ and $b = b_{2}$. In both models $b < 1$ for large values of $\gamma$. Which means that  they have positive pressure at the boundary and are less compact than the Schwarzschild ones, although their  gravitational potential is larger  than in general relativistic models, viz.,  $(\phi \rightarrow 2/3)$. In the Buchdahl limit they impose the same lower limit on $g_{TT}(R_{b})$  as in general relativity, namely, 
$g_{TT}(R_{b}) >  1/9$. At first sight these properties  seem to be  counter-intuitive because one would have thought that models with gravitational potential greater than $4/9$ would be more compact than those in ordinary general relativity. However, this is not so; models with $b <1$ are less compact than in general relativity regardless of $\phi$ or the equation of state at the center.

In terms of $Z = 1/\sqrt{g_{TT}(R_{b})} - 1$, the redshift of the light emitted from the boundary surface,  both the general relativistic Schwarzschild  exterior as well as the braneworld spatial Schwarzschild exterior lead to the  same upper bound, namely
\begin{equation}
Z < 2.
\end{equation}
However, when the external
spacetime is the temporal Schwarzschild metric or the Reissner-Nordstr{\"o}m-like exterior 
there is no such constraint: $Z < \infty$. This infinite difference in the limiting
value of $Z$ is because for these exteriors the pressure at the surface is negative. It is interesting to mention that in Kaluza-Klein gravity the maximum surface redshift is $Z_{KK} = 2.478$ \cite{JPdeLgr-qc/0701129v2}.

Finally, our work demonstrates that non-Schwarzschild exteriors are plenty of new physics,  and are more complicated that in general relativity, both because of the number of technical details, and because of the possible new physical models.

\bigskip

\begin{center}
{\bf{Captions}}
\end{center}

\paragraph{Caption to Figure 1:}  The figure shows that, for large values of $\gamma$, the gravitational potential of stars with $b < 1$ $(b > 1)$ is above  (bellow) the Schwarzschild gravitational potential. For small values of $\gamma$ the situation is the opposite. Namely, the gravitational potential in models with $b > 1$ $(b < 1)$ is higher (lower)  than in the Schwarzschild one. Intersection points indicate that for a given $\phi$ and $\gamma$ there are two possible models;  for  $b = b_{1}$ and $b = b_{2} > b_{1}$. With the increase of $\gamma$, the gravitational potential very rapidly approaches its limiting value $\phi = 2(1 - b/3)/3$. Both models have similar behavior for small values of $b$. 

\medskip

\paragraph{Caption to Figure 2:} $3D$ plot for $b_{1}$.  It illustrates that $b_{1}$ rapidly goes to zero with the increase of $\gamma$, for {\it every} value of  $\phi$. It also shows that $b_{1}$ increases with $\phi$, for every fixed $\gamma$. This increase is sharper for small $\gamma$ than for large $\gamma$.

\medskip

\paragraph{Caption to Figure 3:} $3D$ plot of $b_{2}$. It illustrates that $b_{2}$ is practically insensitive to the change in $\gamma$, for every fixed $\phi$. Also, $b_{2}$ decreases ``almost" linearly with the increase of $\phi$, for {\it all} values of $\gamma$. In particular,  $b_{2} = 3(1 - 3\phi/2)$ in the limit $\gamma \rightarrow \infty$.


\begin{thebibliography}{99}
\bibitem{Maartens1}{R. Maartens, {\em Phys. Rev.} {\bf D62}, 084023 (2000); hep-th/0004166.}
\bibitem{Maartens2}{Roy Maartens, Frames and Gravitomagnetism, ed. J Pascual-Sanchez et al. (World Sci., 2001), p93-119; gr-qc/0101059.}




\bibitem{Dadhich1}{Naresh Dadhich and S.G. Gosh, {\em Phys. Lett.} {\bf B518}, 1(2001); hep-th/0101019.}
\bibitem{Govender}{M. Govender and N. Dadhich, {\em Phys.Lett.} {\bf B538}, 233(2002);  hep-th/0109086.}
\bibitem{Cristiano}{C. Germani and Roy Maartens, {\em Phys. Rev.} {\bf D64},  124010(2001); 
hep-th/0107011.}
\bibitem{Bruni}{M. Bruni, C. Germani and R. Maartens, {\em Phys. Rev. Lett.}
{\bf 87}, 231302(2001);   gr-qc/0108013.}

\bibitem{Kofinas}{G. Kofinas and E. Papantonopoulos, {\em J. Cosmol. Astropart. Phys.} {\bf 12},  11(2004); gr-qc/0401047.}

\bibitem{Wesson 1}{P.S. Wesson, {\em G. Rel. Gravit.} {\bf 16}, 193(1984).}
\bibitem{JPdeL 1}{J. Ponce de Leon, {\em Gen. Rel. Grav.} {\bf 20}, 539(1988).}
\bibitem{Wesson and JPdeL}{P.S. Wesson and J. Ponce de Leon, {\em J. Math. Phys.} {\bf 33}, 3883(1992).}
\bibitem{Coley1}{A.A. Coley and D.J. McManus, {\em J. Math. Phys.} {\bf 36}, 335(1995).}
\bibitem{Overduin}{J.M. Overduin and P.S. Wesson, {\em Phys. Reports} {\bf 283}, 303(1997).}
\bibitem{Coley2}{A.P. Billiard and A.A. Coley, {\em Mod. Phys. Lett.} {\bf A12}, 2121(1997).}
\bibitem{Wesson book}{P.S. Wesson, {\em Space-Time-Matter} (World Scientific Publishing Co. Pte. Ltd. 1999).}
\bibitem{JPdeLgr-qc/0105120v2}{J. Ponce de Leon, {\em Int.J.Mod.Phys.} {D11},  1355(2002);  gr-qc/0105120.}

\bibitem{JPdeLgr-qc/0111011}{J. Ponce de Leon, {\em Mod.Phys.Lett.} {\bf A16}; gr-qc/0111011.}
\bibitem{JPdeLgr-qc/0512067}{J. Ponce de Leon, {\em Class.Quant.Grav.} {\bf 23},  3043(2006); gr-qc/0512067.}
\bibitem{SanjeevWesson}{S.S. Seahra and P.S. Wesson, {\em Class.Quant.Grav.} {\bf 20} 1321(2003); gr-qc/0302015.}
\bibitem{Indefenceof}{P.S. Wesson, ``In Defense of Campbell's Theorem as a Frame for New Physics"; gr-qc/0507107.}

\bibitem{Randall2}{L. Randall and R. Sundrum, {\em Phys. Rev. Lett. } {\bf 83}, 4690(1999); hep-th/9906064.}
\bibitem{Shiromizu}{T. Shiromizu, Kei-ichi Maeda and Misao Sasaki, {\em Phys. Rev.} {\bf D62}, 02412(2000); gr-qc/9910076.}
\bibitem{JPdeLgr-qc/0511067}{J. Ponce de Leon, {\em Mod. Phys. Lett.} {\bf A21}, 947(2006); gr-qc/0511067.}
\bibitem{Dadhich}{N. Dadhich, R. Maartens, P. Papadopoulos and V. Rezania, {\em Phys.Lett.} {\bf B487},  1(2000); 

hep-th/0003061.}
\bibitem{Casadio}{R. Casadio, A. Fabbri and L. Mazzacurati, {\em Phys.Rev.} {\bf D65}, 084040(2002);  
gr-qc/0111072.}
\bibitem{Viser}{M. Visser and D. L. Wiltshire, {\em Phys.Rev.} {\bf D67}, 104004(2003); hep-th/0212333.}
\bibitem{Bronnikov}{K.A. Bronnikov, H. Dehnen and V.N. Melnikov, {\em Phys.Rev.} {\bf D68},   024025(2003); gr-qc/0304068.}
\bibitem{Weinberg}{Steven Weinberg, {\em Gravitation and Cosmology} (John Wiley and Sons, Inc. 1972).}
\bibitem{Bowers}{L. Bowers and E.P.T. Liang, {\em Astrophys. J.} {\bf 188}, 657(1974).}
\bibitem{Old JPdeL}{J. Ponce de Leon, {\em Phys. Rev.} {\bf D37}, 309(1988).}
\bibitem{Buchdahl}{H.A. Buchdahl, {\em Phys. Rev.} {\bf 116}, 1027(1959).}
\bibitem{JPdeLgr-qc/0212058}{J. Ponce de Leon, {\em Gen.Rel.Grav.} {\bf 36},  923(2004); gr-qc/0212058.}
\bibitem{Bronnikov2}{K.A. Bronnikov and S-W Kim, {\em Phys.Rev.} {\bf D67},  064027(2003); gr-qc/0212112.}
\bibitem{JPdeLgr-qc/0701129v2}{J. Ponce de Leon, {\em Class.Quant.Grav.} {\bf 24}, 1755(2007); gr-qc/0701129.}

\end{thebibliography}
\end{document}